\documentstyle[11pt,newpasp,twoside]{article}
\begin{document}

\title{On the inverse Compton scattering model of radio pulsars}

\author{G. J. Qiao$^{1,2}$,
        R. X. Xu$^{1,2}$,
        J. F. Liu$^{1,2}$,
        J. L. Han$^{1,3}$,
        B. Zhang$^{1,2}$}

\affil{
$^1$Beijing Astrophysics Center, CAS-PKU, Beijing 100871,
China\\
$^2$Astronomy Department, Peking University, Beijing
100871, China\\
$^3$National Astronomical Observatories, CAS,
Beijing 100012, China
}

\begin{abstract}

Some characteristics of the inverse Compton scattering (ICS) model
are reviewed. At least the following properties of radio pulsars
can be reproduced in the model: core or central emission beam, one
or two hollow emission cones, different emission heights of these
components, diverse pulse profiles at various frequencies, linear
and circular polarization features of core and cones.

\end{abstract}

\section{Introduction}

``The crisis today, over twenty years later, is that theory has
produced nothing more useful to compare observations with in order
to interpret them'' (Radhakrishnan 1992). On observational aspect,
emission beam of a radio pulsar can be divided into two (core,
inner conical) or three (plus an outer conical) components through
careful studies of the observed pulse profiles and polarization
characteristics (Rankin 1983; Lyne \& Manchester 1988).

Many pulsar profiles at meter wavelength are dominated by core
components. In the usual models of radio pulsars which are related
to polar cap, it is difficult to get a central or ``core''
emission beam. Considering such obvious discrepancy between
observations and theory, several authors (e.g., Beskin et al.
1988; Qiao 1988a,b; Wang et al. 1988, Lyutikov et al. 1999)
presented their models for the core emission. In the following, we
discuss some characteristics of the inverse Compton scattering
(ICS) model (Qiao 1992; Qiao and Lin 1998, hereafter as QL98; Liu
et al. 1999; Qiao et al. 1999; Xu et al. 2000).

\section{On Emission beams}

\subsection{Basic idea of the ICS model}

The basic idea of the ICS model (Qiao 1988a,b; QL98) can be
briefly described as following. Low-frequency electromagnetic
waves are produced in the polar cap due to gap sparking and
afterwards propagate out freely; Such low energy photons are then
inverse-Compton-scattered by the secondary particles from the pair
cascades, and the up-scattered waves are the radio emission
observed from pulsars.

With the conditions $B\ll B_q=4.414\times 10^{13}$ Guass and
 $\gamma \hbar \omega _0\ll m_ec^2$, the frequency formula of
the ICS mechanism (Qiao 1988a,b; QL98) is $$ \omega \simeq 2\gamma
^2\omega_0(1-\beta \cos \theta _i),\eqno(1) $$ where $\omega_0$
and $\omega$ are the frequencies of incident and scattered waves,
respectively. The Lorentz factor of the secondary particles is
$\gamma =1/\sqrt{1-\beta ^2}$, and $\theta _i$ is the incident
angle (the angle between the direction of motion of a particle and
an incoming photon). Using the formula above, we have obtained so
called ``beam-frequency figure'' (see QL98, Qiao et al. 1999),
which is the plot of the beaming angle $\theta _\mu $ (the angle
between the emission direction and the magnetic axis) versus
observing frequency $\nu$. The shape of emission beams, emission
heights and other emission properties can be derived from the
``beam-frequency figure''.

\subsection{The central emission beam and hollow cones}

Two distinct types of emission components were identified from
observations, namely, `core' or central emission beam near the
center and one or two hollow cone(s) surrounding the core. The ICS
model can reproduce one core and two cones at the same time. The
emission beam in the model could consist of (1) Core + inner cone
emission, for pulsars with shorter rotational periods; (2) Core +
inner and outer cones, for pulsars with longer periods.
Furthermore, the core emission may be a small hollow cone in fact,
i.e., not be fully filled. This can be identified from some pulsar
profiles through ``Gaussian de-composition'' (Qiao et al. 1999).

In the model, we found that these emission components are emitted
at different heights. The `core' emission is emitted at the place
closed to the surface of neutron star, the `inner cone' at a
higher place, and the `outer cone' at the highest place.

We also found that the sizes of the emission beams should change
with frequency. As observing frequency increases, the
`core'emission beam becomes narrow; the `inner cone' size
increases or at least keeps constant; but the `outer cone' size
decreases. For given magnetic inclination and impact angles, we
can figure out how the shape of pulse profiles vary with frequency
(see Qiao et al.1999). This is the theoretical base for the
classification of radio pulsars.

\subsection{Classification for radio pulsars}

The pulse profile shapes, especially the variation of profiles
with the observing frequencies are very important for
understanding the emission mechanism of pulsars. In the ICS model,
various pulse profiles and its evolution over a wide frequency
range can be well simulated.  As the impact angle gradually
increases, pulsars can be grouped into two types (and further
sub-types) according to the ICS model (Qiao et al. 1999).

\vspace{2mm}
\noindent
{\it Type} I -- Pulsars with only core and
inner conical emissions. Such pulsars usually have shorter periods
(thus have larger polar caps). There are two sub-types.
Type Ia: Pulsars of this type have very small impact angle, and
normally show single pulse profiles at low frequencies, but triple
profiles at high frequencies. Prototype in this form are PSR
B1933+16, PSR B1913+16.
Type Ib: Pulsars of this type have larger impact angle than that
of Type Ia. Though at low frequencies the pulsars also show single
profiles, they will evolve to double profiles at higher
frequencies, since the lines-of-sight only cut the inner conical
beam. An example of this type is PSR B1845-01.

\vspace{2mm}
\noindent
{\it Type} II -- Pulsars with all three
emission beams. Pulsars with average period often have such a
feature.
IIa: Pulsars of this type have five components at most observing
frequencies, since the small impact angle  makes the line-of-sight
cut all the three beams (the core, the inner cone and the outer
cone). One important point is that in the ICS model, the five
pulse components will evolve to three or even one component at
very low frequencies (see QL98, Fig. 6a, line A). Such a feature
have been observed from PSR B1237+25 (Kuzmin et al. 1998).
IIb: For this type, the impact angle is larger than IIa, so that
at higher frequencies, the line-of-sight does not cut the core
beam. Thus the pulsars show three components at low observing
frequencies, but four components when frequency is higher, and
finally just double profiles at highest frequencies. PSR
B2045$-$16 is an good example.
IIc: Pulsars of this type have the largest impact angle, so that
only the outer conical beam is cut by the line-of-sight. The pulse
profiles of this type are double at all observing frequencies,
with the separation between the two peaks decreasing at higher
frequencies. This is just the traditional radius-to-frequency
mapping which has also been calculated in the curvature radiation
(CR) picture. The prototype is PSR B0525+21.  Alternative
situation of this type is that pulsars have single profiles at
most of observing frequencies, but become double components at
very low observing frequencies. An example is PSR B0950+08.

\section{Linear and circular polarization}

Pulsar radio emission is generally found to be linearly polarized
over all longitudes of profiles, some times as high as up to
100\%. The position angle sweep is in an `S' shape, which can be
well understood within the rotating vector model. However,
depolarization and position angle jumping are often found in the
integrate profiles of some pulsars. Considering the retardation
effect due to relative phase shift between pulsar beam components
(the core and conal emission components are retarded from
different heights in ICS model), we find that the phase shift of
beam centers of the different components could cause the
depolarization and position angle jump(s) in integrated profiles
(Xu et al. 1997).

\subsection{Basic idea for polarization in ICS model}

We have calculated the polarization feature of the ICS model. Now
we would like to present one key point about how the circular
polarization is produced in the ICS model.

As many authors (Ruderman \& Sutherland, 1975; QL98) argued, there
may exist the inner gap above the pulsar polar cap. The continuous
gap formation and breakdown provide both low frequency waves with
$\omega_0\sim10^5$ s$^{-1}$ and out-streaming relativistic
particles with $\gamma\sim 10^3 $. The low frequency waves will be
up-scattered by relativistic particles (moving direction ${\bf
n}_{\rm e}$) to observed radio waves. The frequency ($\omega$) of
out-going waves is determined by eq.(1), while their complex
amplitudes ${\bf E}$ are determined by (Liu et al. 1999)
$$ {\bf E}=C {\sin\theta^\prime \over
\gamma^2(1-\beta\cos\theta^\prime)^2}
     e^{{\it i}({\omega_0 \over {\it c}}R-{\omega\over {\it c}}
{\bf R} \cdot {\bf n}+\phi_0)} {\bf e}_{\rm s}, \eqno {(2)} $$
here {\bf n} is the observing direction, {\bf R}  is the low
frequency wave vector, $\theta^\prime$ is the angle between ${\bf
n}_{\rm e}$ and {\bf n}, $C$ is a constant, {\bf e}$_{\rm s}$ is a
electric unit-vector in the co-plan of {\bf n}$_{\rm e}$ and {\bf
n}, $\phi_0$ is the initial phase determined by incident wave.

Moving out and losing energy via inverse Compton scattering, the
particle undergoes a decay in $\gamma$, which is assumed to be
(QL98)
$$ \gamma=\gamma_0[1-\xi(r-R_0)/R_e]. \eqno {(3)} $$
The electric field of the total scattered wave at a direction {\bf
n} is the sum of {\bf E} of each electron if such electrons are
scattered coherently.

The emission region for a certain $\omega$ can be obtained by
eq.(1). As pointed out in QL98, there are generally three possible
emission zones, corresponding to core, inner and outer cones. The
scattered electromagnetic waves can be superposed coherently if
the low-frequency waves are from {\it same} sparking and the
emission region is {\it smaller} than $2\pi c/\omega_0$. The
coherent superposition of scattered waves will result in circular
polarization as well as linear polarization (Xu et al. 2000).

\subsection{Further considerations and numerical results}

\noindent {\it `Subpulse' circular polarization patterns}~

When the line of sight sweeps across a mini-beam, we can see a
{\it transient} ``subpulse''. If the line of sight sweeps across
the center of a  core or inner conal mini-beam, the circular
polarization will experience a central sense reversal, or else it
will be dominated by one sense, either left hand or right hand
according to its traversal relative to the mini-beam.

\vspace{2mm}

\noindent {\it Subpulse Position Angles}

Subpulse position angles show diverse values, which are generally
centered at the projection of the magnetic field. The variation
range is quite small for sub-pulses from outer cone emission
zones, and becomes larger when emission height decreases, that is,
larger for inner cone and core component. The position angles
values are scattered around the value of the projection of the
magnetic field lines. When emission of all the sub-pulses are
summed up, the mean position angle will have the mean value. So
naturally the mean position angle is related to the projection of
magnetic field lines, exactly as assumed by the rotation vector
model.

\vspace{2mm}

\noindent {\it Circular and linear polarization of mean profiles}

An observer can see bunches of particles all around his line of
sight. The polarization he receives from individual bunches is
different. When the magnetic axis is inclined from the rotation
axis, the probability of the sparkings was assumed to decrease
exponentially with the azimuthal angle from the projection of the
rotational axis in our simulation. We then get significant (but
{\it smaller} than in subpulse) circular polarization for core
components of mean pulses. But this is not the case for the cone
components.

\section{Conclusion}

The ICS model can reproduce many observational properties of radio
pulsars. There are emission beams of core, inner and outer cones.
These beams are emitted from different heights. The pulse profiles
changes with frequencies, similar to what observed from many
pulsars. The transient `sub-pulses' have very high circular
polarization (sometime as strong as 100\%). In mean pulse profile,
circular polarization is much higher in core than that in inner
cone and oute cone.

\begin{acknowledgments}
This work is supported by National Nature Sciences Foundation of
China, by the Climbing project of China, the Doctoral Program
Foundation of Institution of Higher Education in China and by the
Youth Foundation of Peking University.
\end{acknowledgments}

\end{document}